# Emulation workbench for position sensitive gaseous scintillation detectors

**L. Pereira**[a,b,*], **L. M. S. Margato**[a], **A. Morozov**[a,b], **V. Solovov**[a] **and F. A. F. Fraga**[a,b]

[a] *LIP – Coimbra*
*Rua Larga, Coimbra, Portugal*
[b] *Department of Physics, University of Coimbra,*
*Rua Larga, Coimbra, Portugal*
*E-mail*: `luis.pereira@coimbra.lip.pt`

ABSTRACT: Position sensitive detectors based on gaseous scintillation proportional counters with Anger-type readout are being used in several research areas such as neutron detection, search for dark matter and neutrinoless double beta decay. Design and optimization of such detectors are complex and time consuming tasks. Simulations, while being a powerful tool, strongly depend on the light transport models and demand accurate knowledge of many parameters, which are often not available. Here we describe an alternative approach based on the experimental evaluation of a detector using an isotropic point-like light source with precisely controllable light emission properties, installed on a 3D positioning system. The results obtained with the developed setup at validation conditions, when the scattered light is strongly suppressed, show good agreement with simulations.

KEYWORDS: Gaseous imaging and tracking detectors; Detector design and construction technologies and materials; Performance of High Energy Physics Detectors

---

[*] Corresponding author

# 1. Introduction

Detectors based on gaseous scintillation proportional counters with Anger type readout [1] are used in various research areas such as neutron detection [2], search for dark matter [3] and neutrinoless double beta decay [4]. Secondary scintillation, generated in this type of detectors for a single event can produce up to ~$10^6$ photons in tens of nanoseconds, allowing to achieve very high spatial resolution and count rate over large detection areas.

Design and optimization of a gaseous Anger camera are complex and time consuming tasks. Simulations, which are typically used in these tasks, strongly depend on the light models and demand accurate knowledge of many parameters. However, these parameters are often unavailable (e.g. scattering coefficients vs. angle of incidence or PMT detection efficiency as a function of both position on photocathode and angle of incidence) while their measurement require significant efforts.

In this paper we present an alternative approach based on experimental evaluation of a detector implementing an isotropic point-like light source with precisely controllable light emission properties, installed on a 3D positioning system. A detailed description of the experimental setup is given here and the results obtained at validation conditions (strongly suppressed scattered light) are presented. The centroid position estimation algorithm was used to investigate the achievable spatial resolution versus the number of emitted photons and the practical spatial resolution versus distance between the light source and the PMT plane. The obtained results were compared with predictions of simulations. The spatial response of individual PMTs to a point light source was experimentally measured and used with success for position reconstruction with the maximum likelihood algorithm [5].

# 2. Experimental set-up and methods

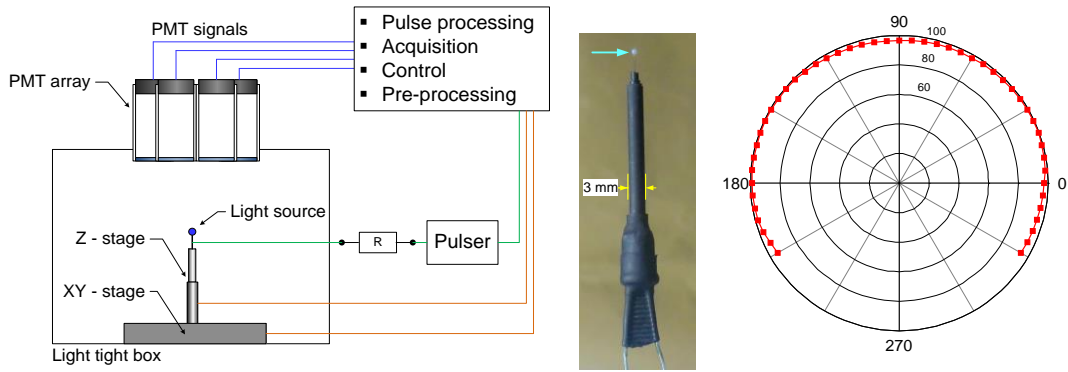

**Figure 1**: Left: Sketch of the experimental set-up. Middle: Photo of a light source (the light blue arrow points to the spherical diffuser). Right: Cross section of the normalized intensity distribution of the diffuser used in this study.

Figure 1 (left) shows a sketch of the experimental set-up. The two main components are a quasi point-like isotropic light source and an array of Hamamatsu R1387 PMTs (38 mm diameter window) placed in a hexagonal configuration with an interaxial distance of 40 mm. The light source is mounted on a 3D positioning system with the accessible volume of $100 \times 100 \times 50$

mm$^3$, and the precision of 25 μm in the XY plane and 250 μm in the Z direction. The light source, the PMT array and the positioning system are assembled inside a light tight box of 500 × 500 × 500 mm$^3$. The large distance between the PMTs and the black painted walls of the box ensure strong reduction of the scattered light reaching the PMTs photocathodes. The source is driven by a Philips PM5786B pulser through a resistor R. The PMT signals were fed to a charge ADC LeCroy 2249W (optionally through a 12 channel amplifier LeCroy 612A for smaller signals) and integrated during an adjustable gate interval. Position control, data acquisition and pre-processing were carried out on a PC.

The light source consisted of a light emitting diode (LED), an optical fiber (OF) and a spherical diffuser made of a photopolymer. The LED (5 mm diameter T1 – ¾) had the brightness of ~6000 mCd and a sharp emission peak at 470 nm. The coupling of the OF to the LED was made through a hole with 0.8 mm diameter and 4 mm length drilled in the epoxy resin encapsulant and fixed with epoxy glue. The diffuser was manufactured with Helioseal (Vivadent, Schaan, Liechtenstein), a dental fissure sealant, in a process similar to the one described in [6]: the optical fiber was immersed in the Helioseal and the LED operated at ~20 mA direct current. At these conditions a 1 mm diameter spherical diffuser could be grown within ~1 minute. After a rinse in methanol the diffuser was hard cured while immersed in paraffin for about 1 hour with the LED operated at the same current. The LED and OF were wrapped in several layers of a black heatshrink tube which provided an adequate light tight, low reflectance enclosure while simultaneously adding mechanical stability. Several diffusers were manufactured and the ones larger in diameter than ~ 1mm and with isotropy (measured with a goniophotometer) worse than 10% in 3π solid angle were rejected.

The required number of photons per pulse was set by adjusting both the pulse amplitude (up to a maximum PMT peak anode current of ~1.5 mA) and the pulse width. A very stable light output (< 2% drift over ~ 10 hours) was observed with the light source set to emit a fixed number of photons per pulse (adjustable between ~1×10$^3$ and ~1×10$^6$) at 500 Hz through a relatively large R (of ~10 kOhm) and using pulse durations from ~10 ns to ~10 μs.

**2.1 PMT characterization**

One of the photomultipliers (PMT$_{Calib}$) was calibrated against a NIST calibrated photodiode (PD) AXUV100G (IRD inc., USA) in order to establish the number of emitted photons per pulse. Using the PD current $I_{PD}$, the number of photons emitted per pulse $N_{Photons}$ is given by

$$N_{Photons} = \frac{I_{PD}}{R_{470\,nm} \cdot E_{ph_{470\,nm}} \cdot f_{\Omega_{PD}} \cdot Freq} \qquad (1)$$

with $R_{470nm}$ being the spectral responsivity of the PD at 470 nm, $E_{ph_{470\,nm}}$ the energy of the emitted photons, $f_{\Omega_{FD}}$ the fractional solid angle subtended by the PD window at the light source position and $Freq$ the frequency of the light pulses. The number of emitted photons can also be calculated from a PMT charge signal distribution characterized by the mean $\mu$ and standard deviation $\sigma$:

$$N_{Photons} = \left(\frac{\mu}{\sigma}\right)^2 \cdot \frac{C}{f_{\Omega_{PMT}}} \qquad (2)$$

where $f_{\Omega_{PMT}}$ is the fractional solid angle subtended by the PMT window at the light source position and *C* is a constant depending only on the electronic properties of the PMT.

To determine the constant *C*, the PD was placed one plane with the $PMT_{Calib}$ window. Operating the light source in steady state conditions, the number of emitted photons $N_{photons}$ was established using equation 1 and the value of the measured PD current $I_{PD}$. Under the same illumination conditions the PMT signal distribution was recorded and *C* calculated from equation 2 using the value of $N_{photons}$ obtained from $I_D$.

During the operation of the system the number of emitted photons was establish through equation 2 by measuring the $PMT_{Calib}$ charge signal distribution at a fixed position with the corresponding geometrical factor $f_{\Omega_{PMT}}$.

The single electron response of the PMTs was recorded with weak pulse intensity when the detection probability of a single photon was ~5%. The measurement was performed pulsing the light source at 100 kHz with 50 ns duration pulses. The signal from the PMT was fed first to a Canberra 2005 preamplifier and then to a Canberra 2020 spectroscopy amplifier. The single electron response spectrum of each PMT was recorded using a multichannel analyzer (CANBERRA 35+) gated synchronously with the light source. The gain *G* of each PMT was estimated by integration of the SER above the pedestal. The found gain values ranged from $0.6\times10^6$ to $1.1\times10^6$.

The detection probability of each PMT is related to the output charge *Q* by the following relation:

$$Det.Prob. = \frac{Q}{N_{ph} \times G \times e} \tag{3}$$

where $N_{ph}$ is the number of photons reaching the photocathode, *G* the mean gain of the PMT and *e* is the charge of electron. The detection probability was evaluated from equation 3 using the gains *G* and the measured charge *Q* when the PMTs were illuminated with a known number of photons $N_{ph}$. The values obtained for the detection probability of the PMTs ranged from 0.14 to 0.20, which matches well the estimation based on the data provided by the manufacturer assuming a typical value of 0.9 for the electron collection efficiency.

## 3. Results and discussion

### 3.1 Simulations and data analysis

All simulations and data analysis in this study were performed with the ANTS2 package [7] (a brief description of the package is available in [8]). The scintillation light source was considered point-like and isotropic. The simulation model took into account the geometry of the PMT array, and the measured values of the gains, detection probabilities and single electron response spectra of the individual PMTs.

### 3.2 Position reconstruction using the center of gravity algorithm

The center of gravity method (CoG) was used for position reconstruction of the emulated scintillation events. This is the most often used method for detectors involving Anger-type readout due to its computational simplicity and robustness. The position estimate $(x, y)$ is given by:

$$x = \frac{\sum_i S_i f_i X_i}{\sum_i S_i f_i}, \quad y = \frac{\sum_i S_i f_i Y_i}{\sum_i S_i f_i} \quad (4)$$

where $(X_i, Y_i)$ are the coordinates of the axis of i-th PMT, $S_i$ is the measured signal and $f_i$ are the weights associated with the individual PMTs (for each PMT $f = 1/(G \times Det.\ Prob.)$. Figure 2 (left) shows a CoG reconstruction of simulated and experimental data for the same regular grid of source positions with $5 \times 5$ mm grid spacing at Z = 30 mm for simulated and experimental data. The average number of emitted photons per pulse was set to ~$5.3 \times 10^5$. At each grid node 1000 events were recorded (or simulated).

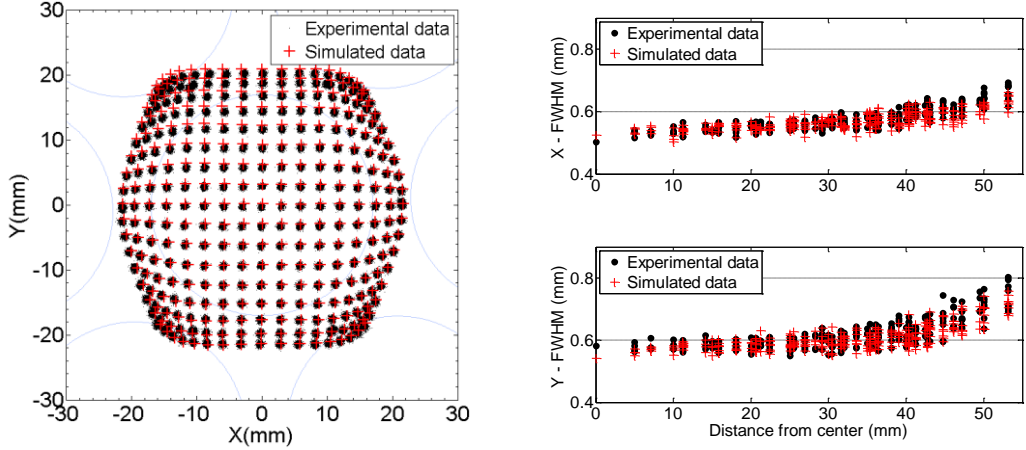

**Figure 2**: Simulation vs experiment. Left: CoG reconstruction of events simulated/recorded for the source sweeping a regular grid of positions. For simulated data only mean reconstructed position is shown for each node. Right: Spatial resolution vs distance from the detector center (0,0).

As it can be seen in figure 2 (left) there is a good agreement between the light source reconstructed for the experimental data and the reconstructed positions of the simulated data. The spatial resolution (FWHM of the reconstructed position) is also very similar for both sets of data (see figure 2 (right)).

### 3.2.1 Spatial resolution vs. number of emitted photons

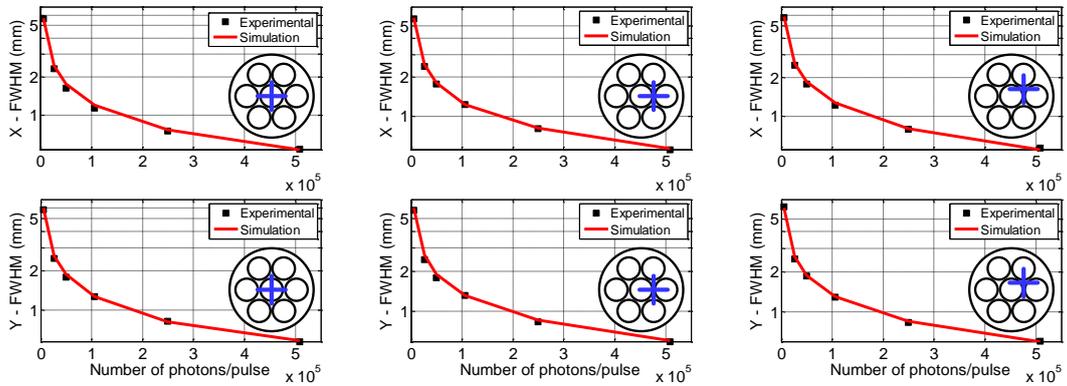

**Figure 3**: Simulation vs experiment: Variation of the spatial resolution with the number of emitted photons per pulse. The position of the light source in the XY plane is shown by the blue cross inside the illustration of the PMT array.

The variation of the spatial resolution of the reconstructed positions versus number of emitted photons per pulse in the range from $5\times10^3$ to $5\times10^5$ is shown in figure 3 measured at Z = 30 mm. The experimental results practically match the simulated ones at these three test locations.

**3.2.2 Spatial resolution vs. distance**

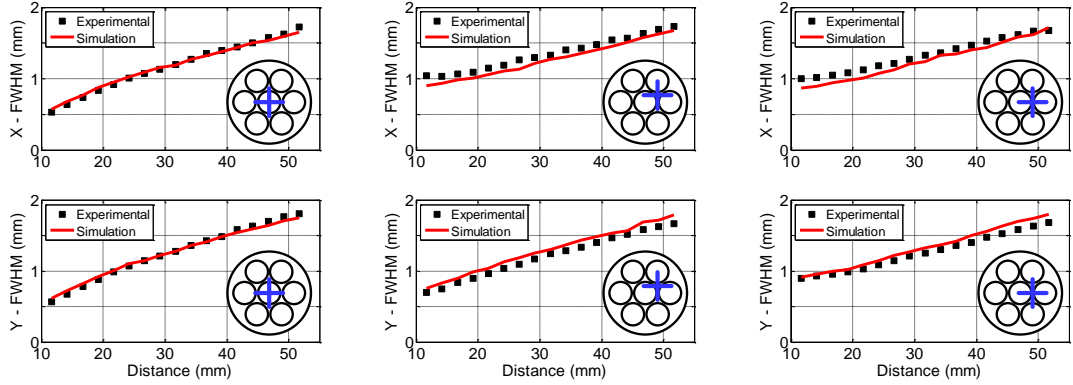

**Figure 4**: Simulation vs experiment: Variation of the spatial resolution on the distance between the light source and the PMT plane. The corresponding light source location in the XY plane is shown by the blue cross inside the illustration of the PMT array.

The variation of the spatial resolution versus the distance between the light source and the PMT plane was measured at three different source positions using the same number of photons per pulse of ~$1\times10^5$. The results are shown in figure 4 demonstrating a very good agreement between the simulated and experimental data.

**3.3 Statistical event reconstruction**

Another reconstruction technique often used for Anger-type detectors involve so-called statistical reconstruction methods. These methods imply an algorithm which searches for the position in which the experimental (or simulated) PMT signals give the best match with the expected PMT signals given by a mathematical model of the detector. For example, the maximum likelihood-based position reconstruction maximize the likelihood function $\mathcal{L}$ [5]:

$$\mathcal{L} = ln\left(\prod_i P_i(n_i)\right) = \sum_i \{n_i ln(N_i(\vec{r})) - N_i(\vec{r})\} + Constant \quad (5)$$

where $P_i$ is the probability for i-th PMT to detect $n_i$ photons when the expected number of photon counts is $N_i$. The calculation is made assuming that the number of detected photons follows the Poisson distribution. The light response functions ($N_i(\vec{r})$ in equation 5, hereafter referred to as LRFs) are defined as the expected number of detected photons versus the distance between the PMT center and the projection of the light source at the PMT plane.

To measure the LRF at a distance Z between the light source and the PMT plane, the light source was sweeping in XY plane through the nodes of a grid of 2.5 mm × 2.5 mm regularly spaced positions and the PMT signal was recorded at each node. The obtained response of the PMTs has nearly perfect axial symmetry (see example in figure 5 (left)), confirming that the scattered light has a small contribution to the signal, as was intended for these verification condition of the setup. The center of symmetry of each PMT, found by fitting a gaussian surface to the corresponding signal mapping had a maximum deviation from the PMT's window axis smaller than 1mm. As it can be seen from figure 5 (center) all PMTs have approximately the same LRF. From figure 5 (right) one can see that the solid angle subtended by the PMT photocathode at the source location, deviates very little from the experimentally measured LRF.

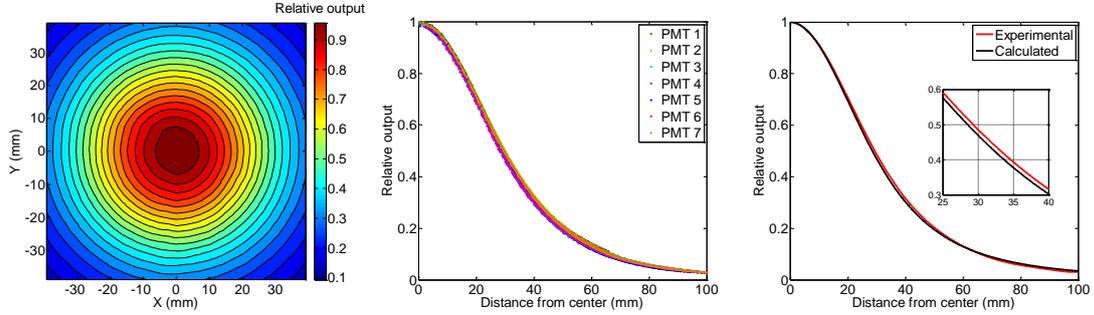

**Figure 5**: Z = 30 mm. Left: Signal of the central PMT versus the light source position. Center: Average PMT signal of a PMT versus distance to the axis of symmetry. Right: Experimental LRF (black) and the calculated LRF considering the diameter of the PMT and the distance between the light source and the PMT plane (red line).

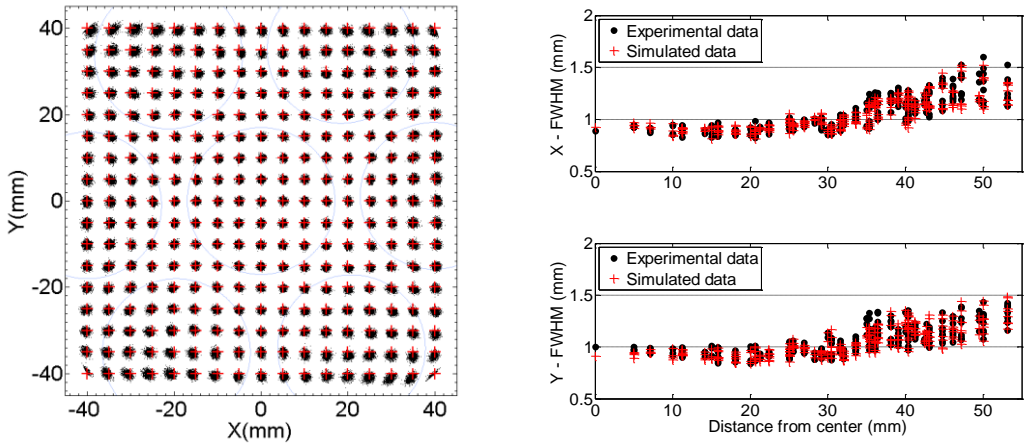

**Figure 6**: Left: Reconstructed positions (black dots) obtained using the maximum likelihood algorithm for experimental data recorded 1000 times at each node of a regular grid of positions, indicated by the red crosses. Right: Spatial resolution (FWHM of the reconstructed position distribution at a grid node) vs distance to the center of the detector (0,0).

Figure 6 (left) shows the ML position reconstruction of the experimental data already described in section 3.2 (black dots). The reconstruction was done using the experimentally measured LRFs (see figure 5 (center)), resulting in a very good reconstruction over the whole active area of the detector (~ 1 PMT diameter from the center). As it can be seen in figure 6 (right) the spatial resolution of the reconstructed experimental data versus distance to the center of the detector is in a very good agreement with the corresponding results obtained for the simulated data (see figure 6 (right)).

## 4. Conclusions

An experimental workbench for emulation of position sensitive gaseous scintillation proportional counters with Anger-type readout has been developed and successfully tested under specific validation conditions with strongly suppressed scattered light. For these conditions we have demonstrated a very good match between the results for the experimental and simulation data in terms of spatial resolution of the detector. We have also demonstrated that for these conditions the approximation on axial symmetry of the PMT spatial response is valid and statistical reconstruction using such LRFs work very well. The workbench can be used to optimize design of detectors after installation of optical elements such as e.g. the decoupling window and the detector walls to properly emulate a real detector with high level of scattered light.

## Acknowledgments

This work was supported by the Portuguese FCT under the grant SFRH / BD / 82505 / 2011.